\documentstyle[12pt,psfig,epsfig,axodraw]{article}
\advance\textheight by \topskip
\begin{document}
\tolerance=100000
\thispagestyle{empty}
\setcounter{page}{0}

\def\cO#1{{\cal{O}}\left(#1\right)}
\newcommand{\be}{\begin{equation}}
\newcommand{\ee}{\end{equation}}
\newcommand{\br}{\begin{eqnarray}}
\newcommand{\er}{\end{eqnarray}}
\newcommand{\ba}{\begin{array}}
\newcommand{\ea}{\end{array}}
\newcommand{\bi}{\begin{itemize}}
\newcommand{\ei}{\end{itemize}}
\newcommand{\bn}{\begin{enumerate}}
\newcommand{\en}{\end{enumerate}}
\newcommand{\bc}{\begin{center}}
\newcommand{\ec}{\end{center}}
\newcommand{\ul}{\underline}
\newcommand{\ol}{\overline}
\newcommand{\ra}{\rightarrow}
\newcommand{\sm}{${\cal {SM}}$}
\newcommand{\as}{\alpha_s}
\newcommand{\aem}{\alpha_{em}}
\newcommand{\ycut}{y_{\mathrm{cut}}}
\newcommand{\susy}{{{SUSY}}}
\newcommand{\Dir}{\kern -6.4pt\Big{/}}
\newcommand{\Dirin}{\kern -10.4pt\Big{/}\kern 4.4pt}
\newcommand{\DDir}{\kern -10.6pt\Big{/}}
\newcommand{\DGir}{\kern -6.0pt\Big{/}}
\def\Ecm{\ifmmode{E_{\mathrm{cm}}}\else{$E_{\mathrm{cm}}$}\fi}
\def\gluino{\ifmmode{\mathaccent"7E g}\else{$\mathaccent"7E g$}\fi}
\def\photino{\ifmmode{\mathaccent"7E \gamma}\else{$\mathaccent"7E \gamma$}\fi}
\def\mgluino{\ifmmode{m_{\mathaccent"7E g}}
             \else{$m_{\mathaccent"7E g}$}\fi}
\def\taugluino{\ifmmode{\tau_{\mathaccent"7E g}}
             \else{$\tau_{\mathaccent"7E g}$}\fi}
\def\mphotino{\ifmmode{m_{\mathaccent"7E \gamma}}
             \else{$m_{\mathaccent"7E \gamma}$}\fi}
\def\ML{\ifmmode{{\mathaccent"7E M}_L}
             \else{${\mathaccent"7E M}_L$}\fi}
\def\MR{\ifmmode{{\mathaccent"7E M}_R}
             \else{${\mathaccent"7E M}_R$}\fi}
\def\lsim{\buildrel{\scriptscriptstyle <}\over{\scriptscriptstyle\sim}}
\def\gsim{\buildrel{\scriptscriptstyle >}\over{\scriptscriptstyle\sim}}
\def\jp #1 #2 #3 {{J.~Phys.} {#1} (#2) #3}
\def\pl #1 #2 #3 {{Phys.~Lett.} {#1} (#2) #3}
\def\np #1 #2 #3 {{Nucl.~Phys.} {#1} (#2) #3}
\def\zp #1 #2 #3 {{Z.~Phys.} {#1} (#2) #3}
\def\pr #1 #2 #3 {{Phys.~Rev.} {#1} (#2) #3}
\def\prep #1 #2 #3 {{Phys.~Rep.} {#1} (#2) #3}
\def\prl #1 #2 #3 {{Phys.~Rev.~Lett.} {#1} (#2) #3}
\def\mpl #1 #2 #3 {{Mod.~Phys.~Lett.} {#1} (#2) #3}
\def\rmp #1 #2 #3 {{Rev. Mod. Phys.} {#1} (#2) #3}
\def\sjnp #1 #2 #3 {{Sov. J. Nucl. Phys.} {#1} (#2) #3}
\def\cpc #1 #2 #3 {{Comp. Phys. Comm.} {#1} (#2) #3}
\def\xx #1 #2 #3 {{#1}, (#2) #3}
\def\NP(#1,#2,#3){Nucl.\ Phys.\ \issue(#1,#2,#3)}
\def\PL(#1,#2,#3){Phys.\ Lett.\ \issue(#1,#2,#3)}
\def\PRD(#1,#2,#3){Phys.\ Rev.\ D \issue(#1,#2,#3)}
\def\preprint{{preprint}}
\def\Ord{\lower .7ex\hbox{$\;\stackrel{\textstyle <}{\sim}\;$}}
\def\OOrd{\lower .7ex\hbox{$\;\stackrel{\textstyle >}{\sim}\;$}}
\def\MCH {$\tilde\chi_1^+$}
\def \CH{{\tilde\chi}^{\pm}}
\def \LSP{\tilde\chi_1^0}
\def \SNU{\tilde{\nu}}
\def \BARSNU{\tilde{\bar{\nu}}}
\def \MLSP{m_{{\tilde\chi_1}^0}}
\def \MCH{m_{{\tilde\chi}^{\pm}}}
\def \MCHMIN {\MCH^{min}}
\def \ET{\not\!\!{E_T}}
\def \LL{\tilde{l}_L}
\def \LR{\tilde{l}_R}
\def \MLL{m_{\tilde{l}_L}}
\def \MLR{m_{\tilde{l}_R}}
\def \MSNU{m_{\tilde{\nu}}}
\def \PROCESS{e^+e^- \rightarrow \tilde{\chi}^+ \tilde{\chi}^- \gamma}
\def \PI{{\pi^{\pm}}}
\def \DM{{\Delta{m}}}
\newcommand{\bQ}{\overline{Q}}
\newcommand{\ad}{\dot{\alpha }}
\newcommand{\bd}{\dot{\beta }}
\newcommand{\dd}{\dot{\delta }}
\def \CH{{\tilde\chi}^{\pm}}
\def \MCH{m_{{\tilde\chi}_1^{\pm}}}
\def \LSP{\tilde\chi_1^0}
\def \MUL{m_{\tilde{u}_L}}
\def \MUR{m_{\tilde{u}_R}}
\def \MDL{m_{\tilde{d}_L}}
\def \MDR{m_{\tilde{d}_R}}
\def \MSNU{m_{\tilde{\nu}}}
\def \MTAUL{m_{\tilde{\tau}_L}}
\def \MTAUR{m_{\tilde{\tau}_R}}
\def \mhf{m_{1/2}}
\def \MST{m_{\tilde t_1}}
\def \CHM{H^\pm}
\def \RPVC{\lambda'}
\def\tth{\tilde{t}\tilde{t}h}
\def\qqh{\tilde{q}_i \tilde{q}_i h}
\def\t1{\tilde t_1}
\def \ta1{\tilde\tau_1}
\def \MET{E{\!\!\!/}_T}  
\def\lapp{\mathrel{\rlap{\raise.5ex\hbox{$<$}}
                    {\lower.5ex\hbox{$\sim$}}}}
\def\gapp{\mathrel{\rlap{\raise.5ex\hbox{$>$}}
                    {\lower.5ex\hbox{$\sim$}}}}
\begin{flushright}
\end{flushright}
\begin{center}
{\Large \bf
Signature of heavy Charged Higgs Boson at LHC in the 1 and 3 prong Hadronic Tau Decay 
channels
}\\[1.00
cm]
\end{center}
\begin{center}
{\large Monoranjan Guchait$^a$, Ritva Kinnunen$^{b}$  
{and} D. P. Roy$^{c}$}\\[0.3 cm]
{\it 
\vspace{0.2cm}
$^a$Department of High Energy Physics\\
Tata Institute of Fundamental Research\\ 
Homi Bhabha Road, Mumbai-400005, India.\\
\vspace{0.2cm}
$^b$Helsinki Institute of Physics, Helsinki, Finland\\
\vspace{0.2cm}
$^c$Homi Bhabha Centre for Science Education\\
Tata Institute of Fundamental Research\\
V.N. Purav Marg, Mumbai- 400088, India
}
\end{center}

\vspace{2.cm}

\begin{abstract}
{\noindent\normalsize 
}
\end{abstract}
We have done a fast simulation analysis of the $\CHM$ signal at LHC in the 1 and 3 prong 
hadronic 
$\tau$-jet channels along with the $t \bar t$ background. The $\tau$ polarization was 
effectively used to suppress the background in both the channels. Combining this with 
appropriate cuts on the $p_T$ of the $\tau$-jet,
the missing $E_T$ and the azimuthal angle between them reduces the background below the 
signal level. 
Consequently one gets a viable $\CHM$ signal up to a mass range of 600-700 GeV at moderate 
to large $\tan\beta$.
\vspace{2cm}
\hskip1.0cm
\newpage

\section*{Introduction}
\label{sec_intro}
The Minimal Supersymmetric Standard Model(MSSM) contains a pair of charged Higgs 
bosons $\CHM$ along with three neutral ones. While it may be hard to distinguish 
any of the neutral Higgs bosons from that of the standard model, the $\CHM$ carries 
the unambiguous hallmark of the MSSM Higgs sector. Therefore, it has an important 
role in the search of MSSM Higgs bosons at the Large Hadron collider(LHC). The 
leading order QCD process 
\br
g b \rightarrow t \CHM + h.c 
\label{eq:Hprod}
\er
gives a sizable production cross section for a heavy $\CHM$ at LHC. However its 
dominant decay mode, $\CHM \to t \bar b$, suffers from a large QCD background
\cite{miller}. A more promising signature comes from its leading sub dominant decay mode,
\br
\CHM \to \tau^\pm \nu_\tau ,
\label{eq:sig}
\er
which accounts for a branching fraction of $\gsim$10\% in the moderate to large $\tan\beta$(
$\gsim$10) region. Moreover one can enhance this signal over the SM
background from     
\br
W^\pm \to \tau^\pm \nu_\tau
\label{eq:bg}
\er
by exploiting the opposite polarizations of $\tau$ i.e. $P_\tau$ = +1 and -1 from the 
signal(\ref{eq:sig}) and background (\ref{eq:bg}) respectively. This was shown to give 
a viable signature for a heavy $\CHM$ boson at LHC in its 1-prong hadronic 
$\tau$ decay channel~\cite{dp}. In particular the signal to background ratio was shown to 
be enhanced significantly by requiring the charged prong to carry $>$ 80\%
of the visible $\tau$ jet energy.

\par
The work of ref~\cite{dp} was based on a parton level Monte Carlo simulation for the 
$\CHM \to \tau^\pm \nu_\tau$ signal(\ref{eq:sig}) and the
$W^\pm \to \tau^\pm \nu_\tau$ background(\ref{eq:bg}), followed by a simple hadronic $\tau$ 
decay code via 
\br
\tau \to \pi^\pm\nu(12.5\%), \rho^\pm \nu(26\%), a_1^\pm \nu(15\%).
\label{eq:taudk}
\er 
The $a_1^\pm$ mode contributes half and half to the 1-prong 
($\pi^\pm \pi^0 \pi^0$) and 3-prong ($\pi^\pm \pi^\pm \pi^\mp$) channels. So 
the three mesons of eq.(\ref{eq:taudk}) account for over 90\% of the 1-prong hadronic 
decay branching ratio(BR) of $\tau$ ($\simeq$50\%)~\cite{pdg}. A more exact analysis was done in \cite{ritva}
following the same procedure, where the signal(\ref{eq:sig}) and background(\ref{eq:bg}) 
were simulated using {\tt PYTHIA} Monte Carlo(MC) event generator\cite{pythia}  
along with the fast {\tt CMSJET} package for detector simulation~\cite{cmsjet}. A 
similar analysis
was also done by members of the ATLAS collaboration\cite{atlas}. In this paper
we have investigated the signal(\ref{eq:sig}) and background(~\ref{eq:bg}) in both 
1 and 3 prong hadronic decay channels of $\tau$ along the lines of 
ref.\cite{ritva}.

However, we have used the {\tt TAUOLA} package\cite{tauola} for hadronic $\tau$ 
decay unlike ref.\cite{ritva}, which had used the simple decay code of 
ref.~\cite{dp} via eq.(\ref{eq:taudk}). The two $\tau$ decay programs give very 
similar results. But the {\tt TAUOLA} package is more exact, since it includes the small non 
resonant contribution to hadronic $\tau$ decay. Besides this is the first investigation
of this process including both 1 and 3 prong hadronic decay channels of $\tau$.

\section*{~$\tau$ Polarization}

It is easy to understand the effect of $\tau$ polarization($P_\tau$) on its 1-prong hadronic 
decay via the dominant contributions of eq.(\ref{eq:taudk}). The center 
of mass angular distributions of $\tau$ into $\pi$ or a vector meson v(=$\rho,a_1$)
is simply given in terms of its polarization as 
\br
{1 \over \Gamma_\pi} {d\Gamma_\pi \over d\cos\theta} &=&
{1\over2} (1 + P_\tau \cos\theta) \nonumber \\
{1 \over \Gamma_v} {d\Gamma_{v L,T} \over d\cos\theta} &=&
{{1\over2} m^2_\tau, m^2_v \over m^2_\tau + 2m^2_v} (1 \pm P_\tau
\cos\theta)
\label{eq:tadist}
\er
where L, T denote the longitudinal and transverse polarization states of the 
vector meson. This angle is related to the fraction $x$ of the $\tau$ lab momentum 
carried by the meson, i.e the (visible) $\tau$-jet momentum
via
\br
x = {1\over2} (1 + \cos\theta) + {m^2_{\pi,v} \over 2m^2_\tau} (1 -
\cos\theta).
\label{eq:cth}
\er  
It is clear from eqs.(\ref{eq:tadist}) and (\ref{eq:cth}) that the signal 
($P_\tau$= + 1)
has a harder $\tau$-jet than the background ($P_\tau$= $-$1) for the $\pi$, 
$\rho_L$ and $a_{1L}$ contributions; but it is the opposite for $\rho_T$ and 
$a_{1T}$ contributions. Now the transverse $\rho$ and $a_1$ decays favor even 
sharing of the momentum among the decay pions, while the longitudinal $\rho$
and $a_1$ decays favor uneven distributions, where the charged pion carries 
either very little or most of the momentum. Thus requiring the $\pi^\pm$ to carry
$\gsim$80\% of the $\tau$ jet momentum,
\br
R_1 = \frac{p_{\pi^\pm}}{p_{\tau-jet}}\gsim0.8,
\label{eq:R1pi}
\er    
retains about half the longitudinal $\rho$ along with the pion but very little of the 
transverse contributions. This cut suppresses not only the $W \to \tau\nu$ 
background, but also the fake $\tau$ background to the 1-prong hadronic
decay channel from QCD jets.

\par 
The 3-prong hadronic decay accounts for about 15\% of $\tau-$decay, of which 2/3rd 
(10\%) comes from 
\br
\tau \to \pi^\pm \pi^\pm \pi^\mp \nu
\label{eq:tadk}
\er
without any accompanying $\pi^0$. We shall consider only this 3-prong decay 
channel, which can be separated by either matching the tracker momentum of the $\tau$-jet with
the calorimetric energy
deposit or by a veto on accompanying $\pi^0 \to 2 \gamma $ in the electromagnetic calorimeter(EM).
This effectively suppresses the fake $\tau$ background from QCD jets
while retaining 2/3rd of the genuine $\tau$ events. As mentioned above 3/4th 
of the 3-prong decay(\ref{eq:tadk}) comes from the $a_1$ contribution. Thus one 
can again enhance the $a_{1L}$ contribution by imposing a cut on the fractional 
$\tau$-jet momentum carried by the like-sign pair, i.e 
\br
R_3 = \frac{p_{\pi^\pm \pi^\pm}}{p_{\tau-jet}} \ne 0.2 - 0.8.
\label{eq:R3pi}
\er   
We shall see that this cut favors the ($P_\tau$=$+$1) signal over the ($P_\tau=-$1) background 
significantly even after including the non-resonant contribution to eq.(\ref{eq:tadk}). Note that
the like sign pion pair in the 3-prong ($\pi^\pm\pi^\pm \pi^\mp$) decay of $a_1$ is analogous 
to the
neutral pion pair in its 1-prong($\pi^0\pi^0 \pi^\pm$) decay, i.e. $R_3$ corresponds to 1-$R_1$
for the $a_1$ channel.  
 
\section*{Signal and Background}
We have computed the $\CHM$ signal from the leading order QCD process(\ref{eq:Hprod}) 
using {\tt PYTHIA}\cite{pythia}. For simplicity we have used a common renormalization and 
factorization scale $\mu_R = \mu_F=\hat s$. The resulting $\CHM$ cross section has 
been enhanced 
by a K-factor of 1.5 to account for the higher order corrections following ref~\cite{zhu}.
Note that the $\CHM$ Yukawa coupling of the signal process(~\ref{eq:Hprod}) is estimated 
using the running quark masses $m_t(\mu_R)$ and $m_b(\mu_R)$. Then it is followed by the 
hadronic decay of top, $t \to b q q^\prime$, along with the $\CHM \to \tau^\pm \nu$ 
decay of 
eq.(\ref{eq:sig}). Finally the hadronic decay of $\tau$ is simulated using {\tt TAUOLA}
\cite{tauola}. Thus the final state consists of at least four hard jets, including the 
$\tau$-jet, along with a large missing-$E_T$ ($\ET$). These events are expected to be 
triggered by a multi-jet trigger along with a higher level $\tau$ trigger with high 
trigger 
efficiency of 90\% at CMS~\cite{ritva}
{\footnote{This trigger efficiency falls to 63-64\%
in full simulation\cite{ritva2}. We have not included it in this work, however, 
since it is
based on fast simulation. Its effect will be to rescale the size of signal and background
shown in the tables as well as figures 1-7 by 2/3rd. Correspondingly, the discovery limits of
$\tan\beta$ in fig.8 will increase by 25\%.}}.The jets and the $\ET$ are reconstructed with the 
fast CMS detector response simulation package {\tt CMSJET}~\cite{cmsjet}. The program 
containscthe detector resolution and the main cracks and inefficiencies.

The background was computed from the leading order $t \bar t$ production process
using $\mu_R = \mu_F = \hat s$ and multiplying the resulting cross-section with the appropriate
K-factor 1.3\cite{topk}. This is followed by the hadronic decay of one top, $t\to b q q^\prime$,
while the other decay  via t $\to b W \to b \tau \nu$. We do not consider the other 
contributions 
to the background of eq.(\ref{eq:bg}) coming from W+multijet production, since they can be 
effectively suppressed by the reconstructions of the hadronic top mass and 
b-tagging~\cite{ritva}.
Note however that the same polarization cut suppresses this W+multijet background as much 
as the $t \bar t $ background.

The $\tau$ identification is based on the narrowness of the $\tau$ jet. To implement 
this we 
define a narrow signal cone of size $\Delta R_S=0.1$ and an isolation cone of 
size $\Delta R_I=0.4$ around the calorimetric jet
axis, where $\Delta R$ is defined via the azimuthal angle and the pseudo-rapidity as 
\br
\Delta R = \sqrt{\Delta\phi^2 + \Delta\eta^2} 
\er
We require 1 or 3 charged tracks inside the signal cone, with $|\eta|<2.5$ and $p_T>$3 GeV for 
the  
hardest track, the former corresponding to the pseudo-rapidity coverage of the tracker.
We further require that there are no other charged tracks with $p_T>$1 GeV inside 
the isolation cone
to ensure tracker isolation~\cite{ritva}. This was shown to be adequate to suppress 
the fake 
$\tau$-jet background
to the $W \to \tau \nu$ events of the CDF experiment in the 1-prong channel but not in the 
3-prong channel~\cite{cdf}. In the latter case one has to combine the tracker 
isolation with 
the requirements of narrow width and small invariant mass($<$ 1.8 GeV) of the $\tau$ jet 
candidate using the calorimeter information to suppress the fake $\tau$ jet 
background\cite{cdf,sasha,gennai}, 
which is beyond the scope of the present work. Note however that in the high- 
$p_T$ ($\gsim$ 100 GeV)range 
of our interest, the 3-prong 
$\tau$-jet is expected to be tagged by the secondary vertex\cite{gennai}.
Moreover, we are interested in a sub sample of 3-prong $\tau$-jets, without accompanying 
$\pi^0$s, for which the fake background is 
relatively small, as we shall see below.

We ensure the absence of $\pi^0$s by requiring the energy of the 3 charged tracks 
measured in the 
tracker to match with the calorimetric energy deposit of the $\tau$-jet within the 
calorimetric 
energy resolution, i.e
\br 
\Delta E = |E_{trk}^{tot} - E_{cal}^{tot}| < 10 GeV. 
\label{eq:de}
\er
With this cut the $\tau$-jet identified by the tracker matches well with the 
actual $\tau$-jet of the event generator. Therefore, we shall use the tracker 
identification of $\tau$-jet in both 1 and 3 prong channels.

In our simulation, as a basic sets of selection cuts for jet reconstruction, we apply 
\br
p_T^j > 20~GeV, |\eta^j| <4.5
\label{eq:ptcut}
\er   
for all jets, the latter corresponding to the pseudo-rapidity coverage of the 
calorimeter. We require the minimum separation of 
\br
\Delta R > 0.5
\label{eq:dr}
\er 
between jets. The missing-$E_T$ arising mainly due to the presence 
of $\nu_{\tau}$ accompanied with $\tau$ lepton, is reconstructed using the calorimetric
informations, and we set a minimum cut
\br
\MET > 30~GeV.  
\label{eq:met}
\er
The $\MET$ is expected to be large for signal as it originates mainly from the
massive $\CHM \to \tau \nu$ decay, where as in the case of background it comes from 
a relatively light $W \to \tau \nu$ decay.

In Fig.1 we show the distribution of $p_T$ of $\tau$-jet($p_T^{\tau-jet}$) for signal 
in the 
upper panel 
along with the background from $t \bar t$ in the lower panel.  
These distributions are subject to only the basic selection cuts of 
eqs.(\ref{eq:de}-\ref{eq:met}) and 
are normalized for integrated luminosity ${\cal L} = 100 fb^{-1}$. The signals are 
shown for 
two masses of $\CHM$, $m_{\CHM}=$300 and 600~GeV,
for $\tan\beta=40$. Evidently, the higher the mass of $\CHM$, the harder are 
the $\tau$-jets.
Notice that the signal cross section
is several orders of magnitude less than the background even for $p_T^{\tau-jet}>$100 GeV.
  
In Fig.2 we demonstrate the distribution in $R_{1}$(~\ref{eq:R1pi}) for 
1-prong decay
channel of $\tau$-jet for both signal and background with $p_T^{\tau-jet}>$ 100 GeV. 
The spillover of the distribution to the $R_1>$1 region reflects the 15-20\% uncertainty 
in the calorimetric
energy measurement of the $\tau$-jet. As discussed above, for the signal process where 
$P_\tau$ = $+$1, 
peaks occur at the two ends because of the uneven sharing of energy between pions
while in the
case of background, for which $P_\tau$ = $-$1, a peak occurs at the 
middle due to the almost equal sharing of energy between pions. 
Note that at the large $R_1$ end of the distributions 
the dominant contribution comes from $\rho_L$ and $\pi$ channels. 
Therefore a    
cut like $R_{1}>$0.8 leads to a very large suppression of the background while 
retaining almost half
of the signal events. Note that one would in any case 
require a  reasonably hard charged track, corresponding to $R_1\gsim$0.3, for 
effective $\tau$ identification
and rejection of the QCD jet background~\cite{sasha}. So extending this cut to 
$R_1 >$0.8 costs very 
little to the signal, while it effectively suppresses the $P_\tau$=$-$1 background.  
For 3-prong decay of $\tau$-jets we present the $R_{3}$ distribution in Fig.3 for both 
signal and background, where $R_{3}$ is defined by eq.(\ref{eq:R3pi}). We have shown the 
distribution for two values of $\CHM$ mass as before. As mentioned above, the like 
sign pair of 
$\pi$s plays the same role as the $\pi^0$ pair in the case of 1-prong decay channel of 
$\tau$-jet via $a_1$ i.e., the $\pi^0\pi^0\pi^\pm$ channel. 
Hence, the like sign pion pair carries either very little or most of the $\tau$-jet 
energy for the signal($P_\tau=+1$), while it carries roughly 2/3rd of the $\tau$-jet 
energy for the background($P_\tau=-1$). 
Consequently selecting events in the region $R_3<$0.2 and $R_3>$0.8 helps to suppress 
the background. We found the efficiency due to this selection cut is 
about 55\% for signal and 35\% for background. Note that there is no constraint on $R_3$ for
$\tau$-identification unlike the 1-prong channel.  

Since the $\CHM$ production(\ref{eq:Hprod}) is accompanied by a top quark, it is 
useful to reconstruct top quark mass. We perform the $W$ and top mass reconstructions 
from hadronic 
decay modes by requiring at least three reconstructed jets in addition to a single 
$\tau$ jet. For the $W$ mass reconstruction we require the invariant mass of two
jets out of them to be 
\br
m^{rec}_{jj} = m_W \pm 15 ~GeV 
\er
and the corresponding $W$ momenta are obtained out of these jet momenta. In case of several pairs 
satisfying this mass band, the pair having invariant mass closest to $m_W$ is chosen.
The top mass 
reconstruction is performed 
using this pair and one of the remaining jets and 
demanding 
\br
m^{rec}_{Wj}=m_t \pm 30 ~GeV. 
\er
Notice here that we have not done any kind of b-tagging. However to take care of 
b-tagging we multiply the signal and background cross section by a b-tagging 
efficiency($\epsilon_b$) factor 0.5~\cite{ritva}. This 
includes the loss of efficiency due to the reduced coverage of $|\eta|<$2.5 for this jet.
Though the $t \bar t$ background has 2 b jets, one still gets efficiency factor 
2$\times 0.5\times$(1-0.5)=0.5 by requiring only one b-tag.

We have also investigated the relative azimuthal opening angle in the transverse plane 
between $\tau$ jet and the $\MET$ vector, which is also connected with the transverse 
mass via,
\br
m_T = \sqrt{2. p_T^{\tau-jet}. \MET ( 1 - \cos{\Delta\phi(\tau-jet, \MET)})}.
\er   
Since, in the signal process(\ref{eq:sig}) both $\tau$-jet and $\MET$ are originating 
from a comparatively massive $\CHM$ particle, leading to harder $\tau$-jet and missing 
momentum, it is expected that the signal 
distributions in $m_T$ and $\Delta\phi(\tau-jet, \MET)$ will be much broader than the 
background(\ref{eq:bg}), 
which can distinguish the signal and background events.   
In Fig.4 we show the distribution in $\Delta\phi(\tau-jet, \MET)$ for the signal 
for two values of $m_H=$300~GeV and 600~GeV and 
$\tan\beta=$40 along with background. These distributions are for 1-prong $\tau$-jet events 
passed by the selection cuts:   
$E_T^{\tau-jet}> $100 GeV, $R_{1}>$0.8, $E{\!\!\!/}_T >$ 100 GeV. 
In Fig.5 we show the same distribution in $\Delta\phi(\tau-jet, \MET)$ for $\tau$-jets 
decaying via 3-prong decay modes.In Fig.6 and Fig.7 we present the distributions in
$m_T$ for 1-prong and 3-prong channels of $\tau$-jet respectively. 
Evidently, as expected the background is concentrated at small azimuthal opening angle 
$\Delta\phi \simeq$0, while the signal is peaked at the largest opening angle 
$\Delta\phi\simeq 180^0$. Likewise the background distribution in $m_T$ is restricted
to $m_T <m_W$, while the signal distribution is peaked at much larger $m_T$ and 
extends all the 
way up to $m_H$, which can also be used to estimate the $\CHM$ mass. Thus either a cut 
on $\Delta\phi$ or $m_T$ can suppress the 
background by enormous amount without practically any loss of signal events. We will 
see later that a 
$\Delta\phi>$60$^0$ cut brings down the background level to less than the signal.

In our simulation we have generated $10^6$ events each for signal and background
processes in the $\tau$+multijet chanel. In Table 1 we demonstrate the cumulative effect of our selection cuts 
for 1-prong decay channel of 
$\tau$ jet, while Table.2 shows the same for 3-prong decay channel of $\tau$ jet. 
We present the number of events surviving out of $10^6$ generated events after each cut.
In both the tables we present the results
for signal for two sets of values of $m_{H^\pm},\tan\beta=$(300~GeV,40) and (600~GeV,40) 
as well as for
$t\bar t$ background.
First row shows the number of hadronically decaying $\tau$ particles which come 
from $\CHM$ decay of
eq.(\ref{eq:sig}) and W decay background of eq.(\ref{eq:bg}). It reflects the 
hadronic $\tau$ decay
branching ratio of $\sim$65\%. The second row 
shows the number of events having an identified $\tau$-jet, which are subject to the 
basic cuts of      
eqs.(\ref{eq:ptcut}-\ref{eq:met}) and where $\tau$-identification is performed using tracker 
informations as discussed above. 
As expected the corresponding efficiencies for signal are higher than the 
background because of harder $\tau$-jet from $\CHM$ decay. 
For the same reason the efficiency after the  
$p^{\tau-jet}_T>100$~GeV cut, shown in the third row, is about 2.8\% for the background 
where as for signal it is 33\%(58\%) for $m_H=300(600)$~GeV. In the signal events about 
3/4 contribution comes from the 1-prong decay channel and 1/4 contribution is from 3-prong 
decay channel of $\tau$, where as for the background it is 60\% from 1-prong channel and 40\% 
from 3-prong decay channel of $\tau$. The ratio of signal events in the 1 and 3 prong channels 
agree with the respective $\tau$ branching fractions of 50 and 15\%. This shows that the 
$\tau$ identification via the tracker as described 
above works quite well for the signal events. On the other hand there is a clear excess of 3-prong
events in the $t \bar t$ background, showing a large contamination of fake $\tau$ from hadronic
jets in this channel, as mentioned 
earlier. However, they are removed after the removal of accompanying $\pi^0$s via the $\Delta E$
cut of eq.(\ref{eq:de}). After this cut the 3-prong events are about 1/5th of the 1-prong events 
for both the signal and background in agreement with the respective $\tau$ branching fractions 
of 50 and 10\%.     

The next rows show the effects of $\tau$ polarizations on the signal and background. 
The $R_1>$0.8 cut
retains about 40\% of the signal against only 10\% of the background. In fact the effective 
loss to the 
signal is quite small, since the low $R_1$ peak of Fig.2 would be lost anyway due to the 
requirement of a hard charged track for $\tau$ identification as mentioned earlier. The 
corresponding cuts of $R_3>$0.8 or $<$0.2 retains
about 55\% of the signal against 35\% of the background. It shows the efficacy of $\tau$ 
polarization in the extraction of the $\CHM$ signal in the 1-prong as well as 3-prong 
$\tau$-jet channels. Note also that the measurement of the polarization variables 
$R_1$ and $R_3$ is quite simple, since it only requires measuring the momenta of the
charged tracks in the tracker. 

The $\MET>$100 GeV cut has an efficiency of 60(90)\% for a $m_{H^\pm}=$300(600)~GeV Higgs signal 
and 40\% for the $t \bar t$ background. This is followed by the $W$ and top mass cuts which 
have a combined efficiency 20\% for the signal as well as background. Its main utility 
is in suppressing the $W$+multijet background, as mentioned earlier. Finally, 
the $\Delta\phi>60^0$
cut suppresses the background by a factor of 50, with very little loss to the signal.

The last but one rows show the signal and background cross-sections in the $\tau$+multijet
channel, representing the $10^6$ generated events. It corresponds to $\sigma_{t \CHM}\times$
Br($\CHM \to \tau \nu)\times$2/3 for the signal and 2 $\sigma_{t\bar t}\times1/9\times 2/3$ 
for the background. The last rows show the signal and background cross-sections remaining 
after all the cuts, which includes a b-tagging efficiency factor of 0.5. We see that at 
this 
stage the background is reduced to less than the signal size. It can be still further 
suppressed via the
transverse mass distribution without any loss of signal. The $m_T$ distribution can also
be used to estimate $m_H$, as mentioned earlier. Thus the discovery limit is primarily controlled
by the size of the signal shown in the last row. Fig.8 shows the discovery limit corresponding to
25 signal events for 1-prong and (1+3)-prong channel of $\tau$-jet. This is shown
separately for ${\cal L}=30 fb^{-1}$ and 100$fb^{-1}$ expected from the low and high 
luminosity runs of LHC separately. It shows a promising discovery potential for $H^\pm$ over the 
mass range upto 600-700 GeV at moderate to large $\tan\beta$. However it should be noted 
here that a full simulation for the high luminosity run has not been completed 
yet\cite{gennai}. It would require higher trigger threshold of $\tau$-jets, which 
would move up the discovery limits at 
low $m_H$ to slightly higher values of $\tan\beta$.
             
\section*{Summary}          
 
We have investigated the $\CHM$ signal at LHC in the 1 and 3 prong hadronic $\tau$-jet 
channels along
with the $t \bar t$ background. The signal and background processes were generated using 
the {\tt PYTHIA}
event generator interfaced with {\tt TAUOLA} for the 1 and 3 prong hadronic $\tau$ decays. 
The detector
response was simulated using the fast simulation packages {\tt CMSJET}. We have shown
that the opposite polarization of $\tau$ from the signal $(\CHM \to \tau \nu)$ and 
background 
$(W^\pm \to \tau \nu)$ processes can be effectively used to suppress the background with 
respect to the 
signal in both 1 and 3-prong $\tau$-jet channels. The signal was also found to have much 
harder 
distributions than the background in the azimuthal opening angle($\Delta\phi$) between
the $\tau$-jet and the missing $E_T$ as well as in the transverse mass ($m_T$) of the two. 
Combining 
these distinctive features with those of $\tau$ polarization we could effectively 
suppress the background to below the signal size. Thus the $\CHM$ discovery potential 
of LHC in the 1 and 3 prong $\tau$-jet channels is primarily determined by the signal 
cross section in these channels. We find a promising $\CHM$ signal at LHC
over the mass range of several hundred GeV at moderate to large $\tan\beta$.      
 
\section*{Acknowledgements}
One of the authors(DPR) acknowledges partial financial support from BRNS(DAE) through the 
Raja Ramanna Fellowship scheme. 

\newpage
{\bf Table 1}:
Number of events after each set of cuts for two sets of ($M_H,\tan\beta$) values for the 
signal 
and the $t\bar t$ background process. Number of $\tau$+multjet events generated in each case is $10^6$.
Last two rows show the production cross sections in the $\tau$+multijet channel and cross
section after multiplying with efficiency factors including the b-tagging efficicency for
the 1-prong hadronic $\tau$ decay channel.
\begin{center}
\begin{tabular}{|c|c|c|c|}
\hline
 Cuts &$m_H, \tan\beta$  & $m_H, \tan\beta$& Bg\\
  &(300,40) & (600,40) & $t \bar t$\\
\hline
No. of had.$\tau$ decay event & 640531 & 641346 & 641009\\
\hline
Identified $tau$ jets & 448288 & 465556 & 229622\\
\hline
$E_T^{\tau-jet}> $100 GeV & 212491 & 370562 & 17875\\
\hline
1 prong decay & 158865 & 281428 & 10182\\
\hline
$R_{1\pi}>$0.8 & 57240 &  103470 & 1046\\
\hline
 $E{\!\!\!/}_T >$ 100 GeV & 32441  & 89587& 410  \\
\hline
Number of jets $\ge$3 & 15083 & 41783 & 308 \\
\hline
W mass rec, $m_{jj}=m_W \pm 15$GeV & 9861 & 27091 & 147 \\
\hline
Top mass rec,$m_{jW}=m_t \pm 30$GeV & 6376 & 17339 & 87  \\
\hline
$\Delta\phi(\tau-jets,E{\!\!\!/}_T)> 60^0 $ & 5154 & 16394 & 2 \\
\hline
$\sigma \times$BR (pb) & 0.431 & 0.045 & 73\\
\hline
Cross section $\times$ efficiency$\times \epsilon_b$(fb) & 1.1  & 0.37  & .15\\
\hline
\end{tabular}
\end{center}

\vspace{1cm}
{\bf Table 2:}
Same as in Table 1, but for $\tau$ jets in 3-prong decay channel. 
\begin{center}
\begin{tabular}{|c|c|c|c|}
\hline
Cuts &$m_H, \tan\beta$  & $m_H, \tan\beta$& Bg\\
  &(300,40) & (600,40) & $t \bar t$\\
\hline
No. of had.$\tau$ decay event & 640531 & 641346 & 641009 \\
\hline
Identified $tau$ jets & 448288 & 465556 & 229622 \\
\hline
$E_T^{\tau-jet}> $100 GeV & 212491 & 370562 & 17875\\
\hline
3 prong decay & 53626 & 89134 & 7683\\
\hline
$\Delta E < 10$ GeV & 32886 & 54901 & 2610 \\
\hline
$R_{3\pi}<$0.4 or $>0.8$ & 18159 & 29714 & 858 \\
\hline
 $E{\!\!\!/}_T >$ 100 GeV & 10456  & 26173 & 269 \\
\hline
Number of jets $\ge$3 & 4854 & 12161 & 206 \\
\hline
W mass rec, $m_{jj}=m_W \pm 15$GeV & 3138 & 7886 & 110 \\
\hline
Top mass rec, $m_{jW}=m_t \pm 30$GeV & 2010 & 5073 & 60  \\
\hline
$\Delta\phi(\tau-jets,E{\!\!\!/}_T)>$  60$^0$ & 1676 & 4881& 1  \\
\hline
$\sigma \times$Br(pb) &.431&0.045 & 73 \\
\hline
Cross section $\times$ efficiency$\times \epsilon_b$ (fb) & 0.36   &0.11  & 0.07\\
\hline
\end{tabular}
\end{center}

\begin{figure}
\vskip-1cm
\includegraphics[width=6.5in, height=8.3in]{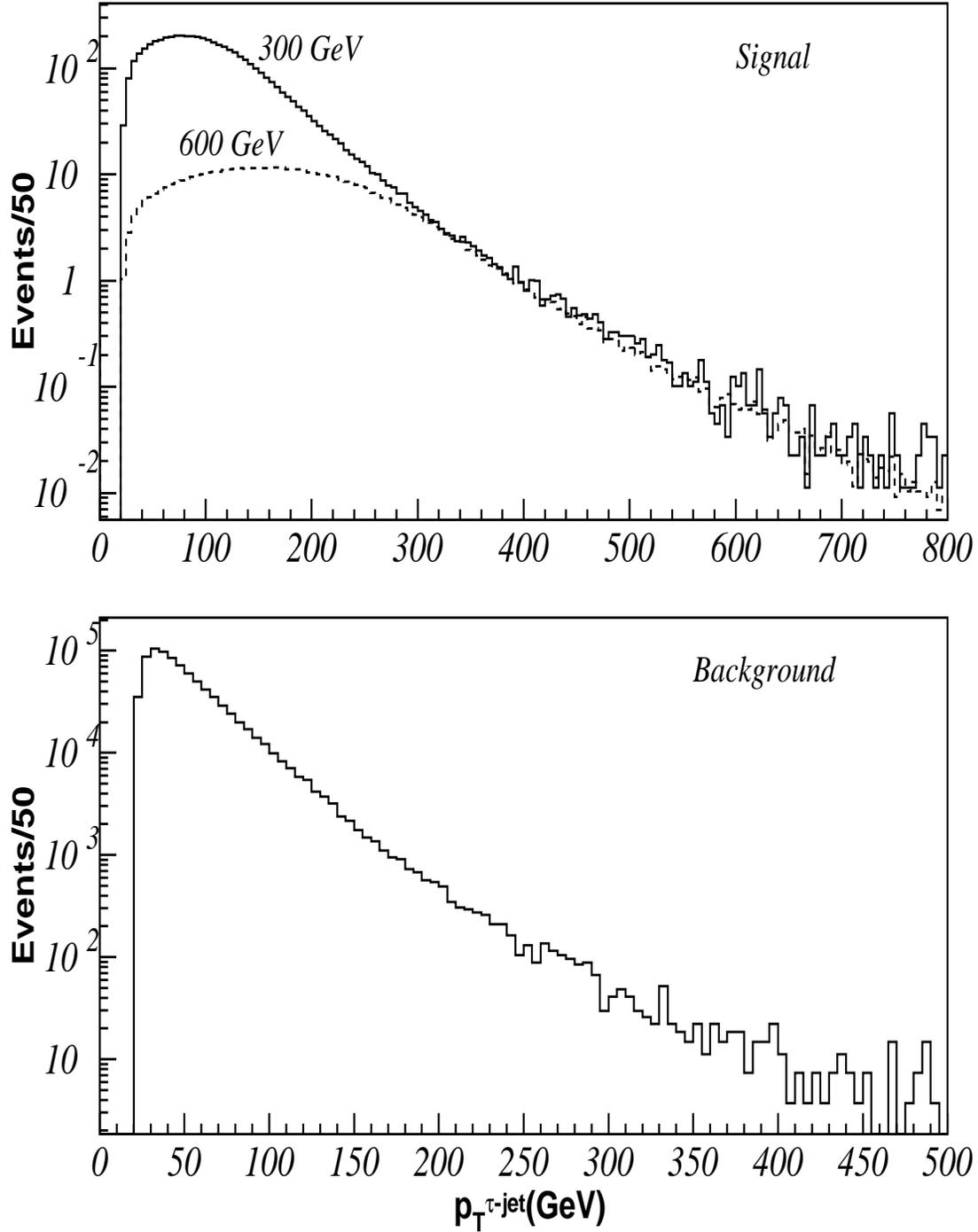}
\vskip-0.7cm
\caption{The number of events are shown against $p_T$ of $\tau$-jets for signal and
background processes for integrated luminosity ${\cal L}=$100$fb^{-1}$ 
and are subject to $p_T^{\tau-jet}> $20 GeV and $|\eta_{\tau-jet}|<$2.5 cuts.
The mass of charged Higgs and 
$\tan\beta$ are set to $m_{H^\pm}=$300 and 600 GeV and $\tan\beta=$40.
}
\end{figure}

\begin{figure}
\vskip-1cm
\includegraphics[width=6.5in, height=8.3in]{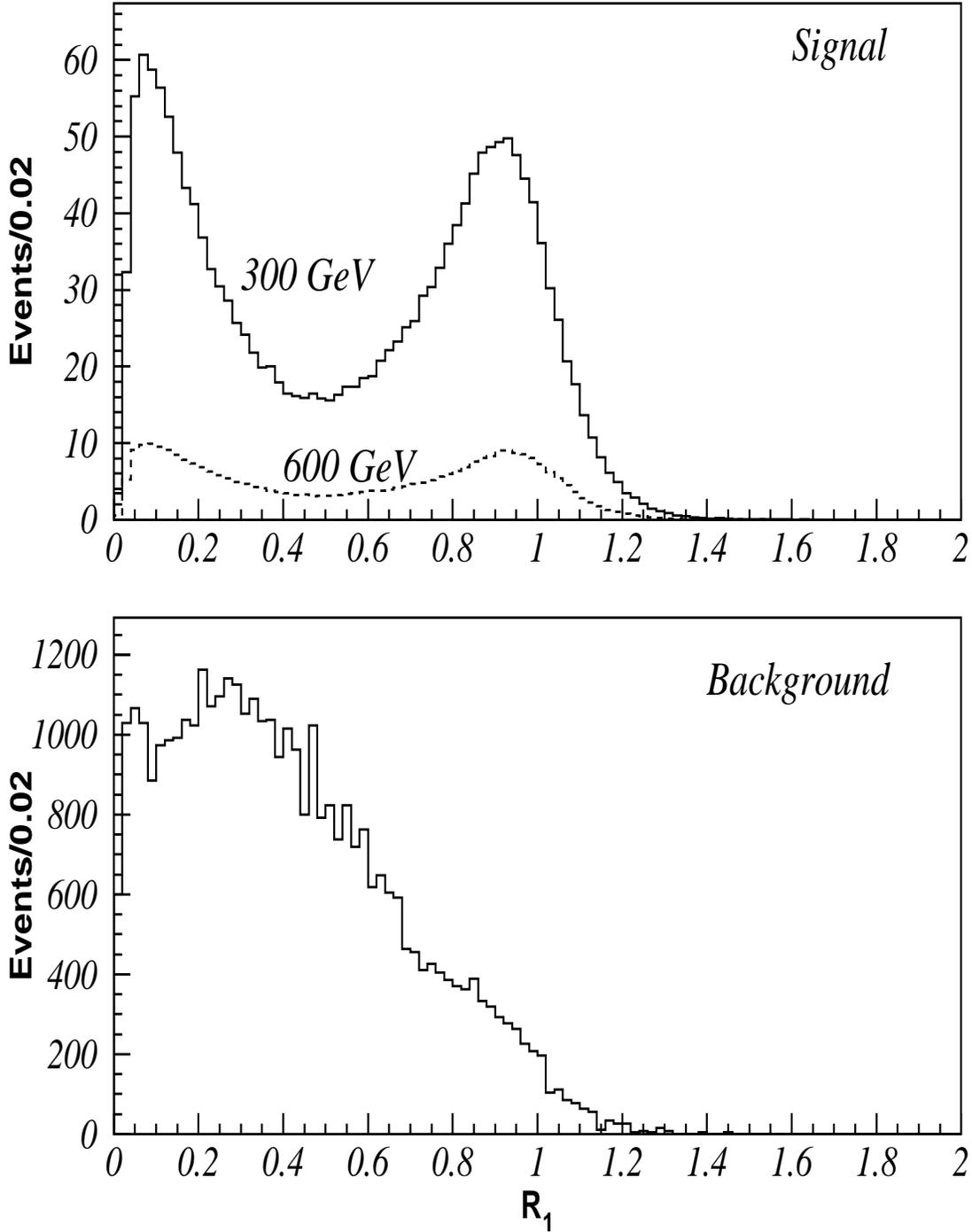}
\vskip-.7cm
\caption{The number of events are plotted against the 
fraction of $\tau$-jet momentum carried by the charged prong $R_1$ for the 1-prong decay
channel of $\tau$-jet for signal and background processes. Both the distributions are 
subject to
$p_T^{\tau-jet}> $100 GeV and $|\eta_{\tau-jet}|<$2.5 cuts.The masses of charged Higgs and 
$\tan\beta$ are same as in Fig.1. 
}
\end{figure}

\begin{figure}
\vskip-1cm
\includegraphics[width=6.5in, height=8.3in]{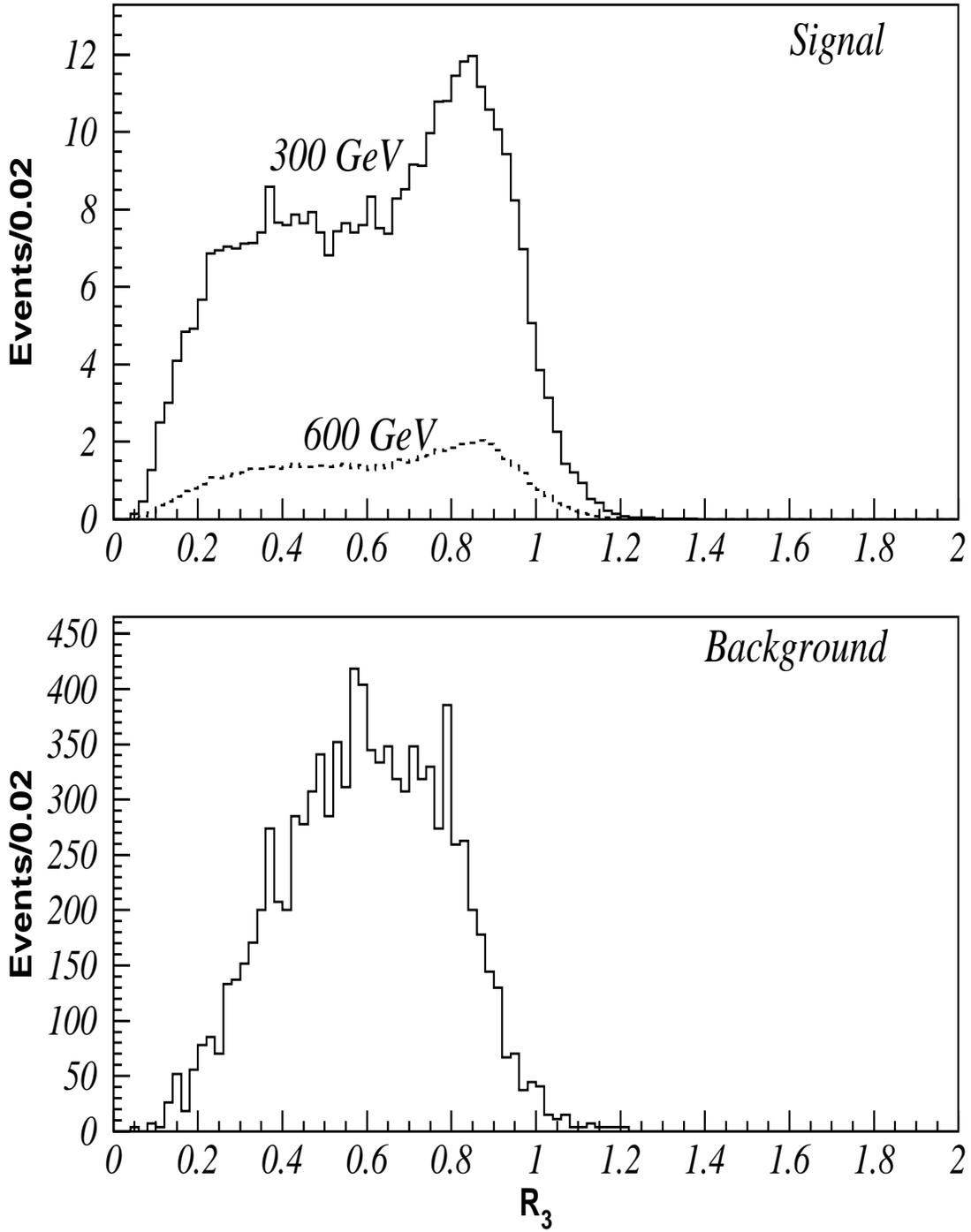}
\vskip-0.5cm
\caption{
Same as in Fig.2, but against the fraction of $\tau$-jet momentum carried by the like sign
pair($R_3$) for the 3-prong $\tau$-jet channel.
}
\end{figure}

\begin{figure}
\vskip-1cm
\includegraphics[width=6.5in, height=8.3in]{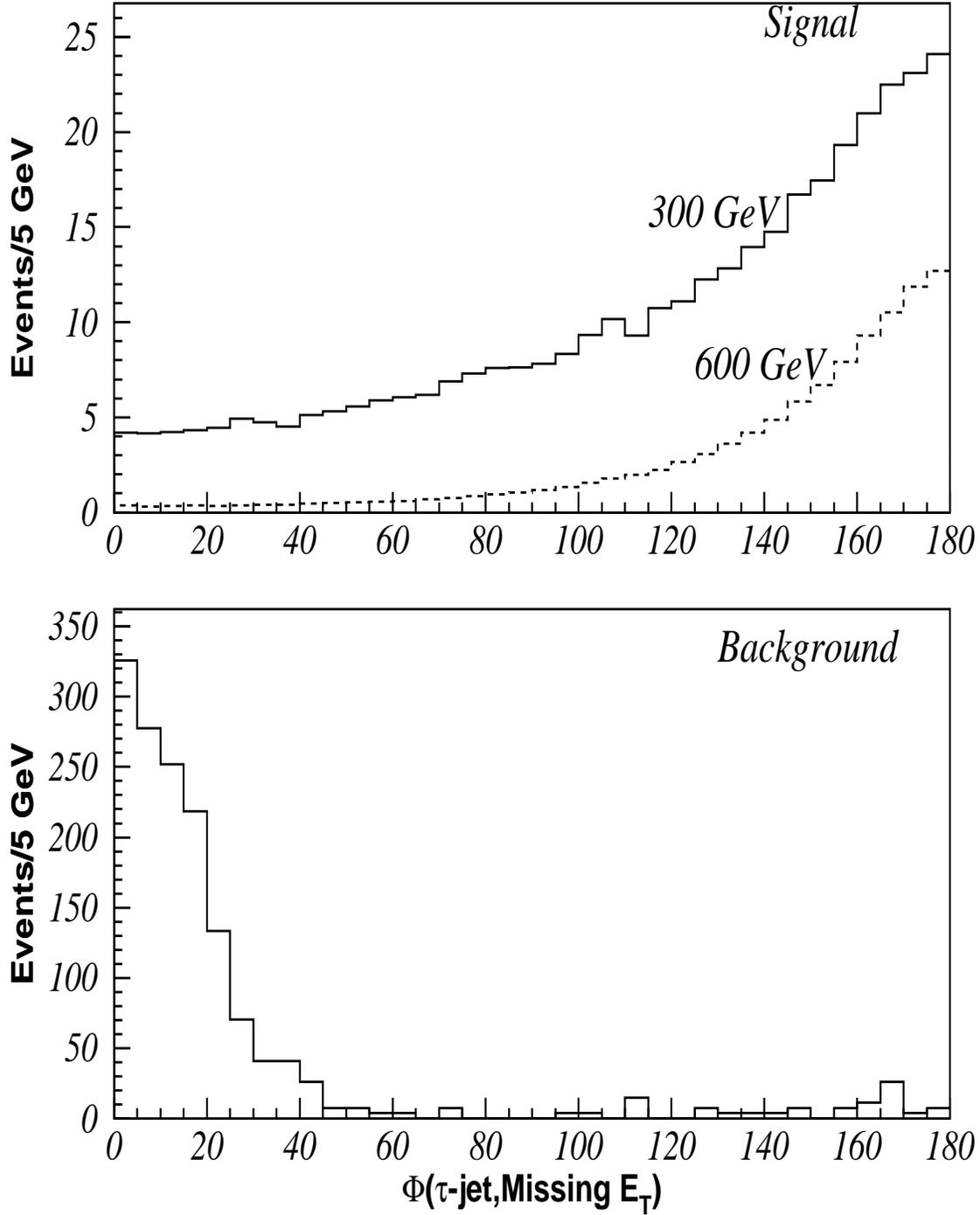}
\vskip-0.5cm
\caption{The number of events are shown against the opening azimuthal angle 
$\Delta\phi(\tau-jets,E{\!\!\!/}_T)$ for signal and background for 1-prong decay 
channel of
$\tau$-jets. These are subject to
$p_T^{\tau-jet}> $100 GeV, $R_{1}>$0.8 and $E{\!\!\!/}_T >$ 100 GeV cuts.
}
\end{figure}

\begin{figure}
\vskip-1cm
\includegraphics[width=6.5in, height=8.3in]{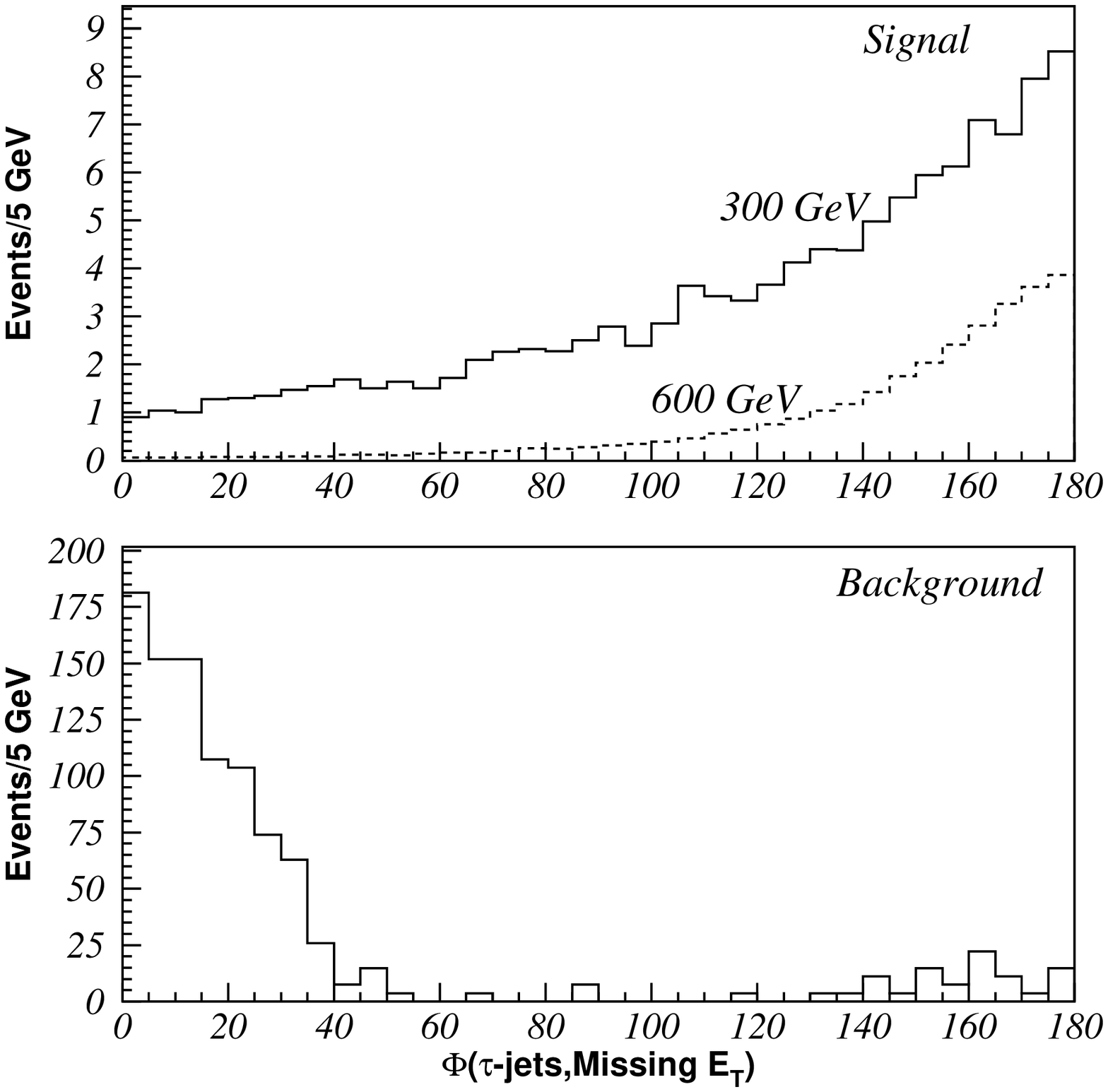}
\vskip-0.7cm
\caption{
Same as in Fig.4, but for 3-prong decay channel of $\tau$-jets.
}
\end{figure}

\begin{figure}
\vskip-1cm
\includegraphics[width=6.5in, height=8.3in]{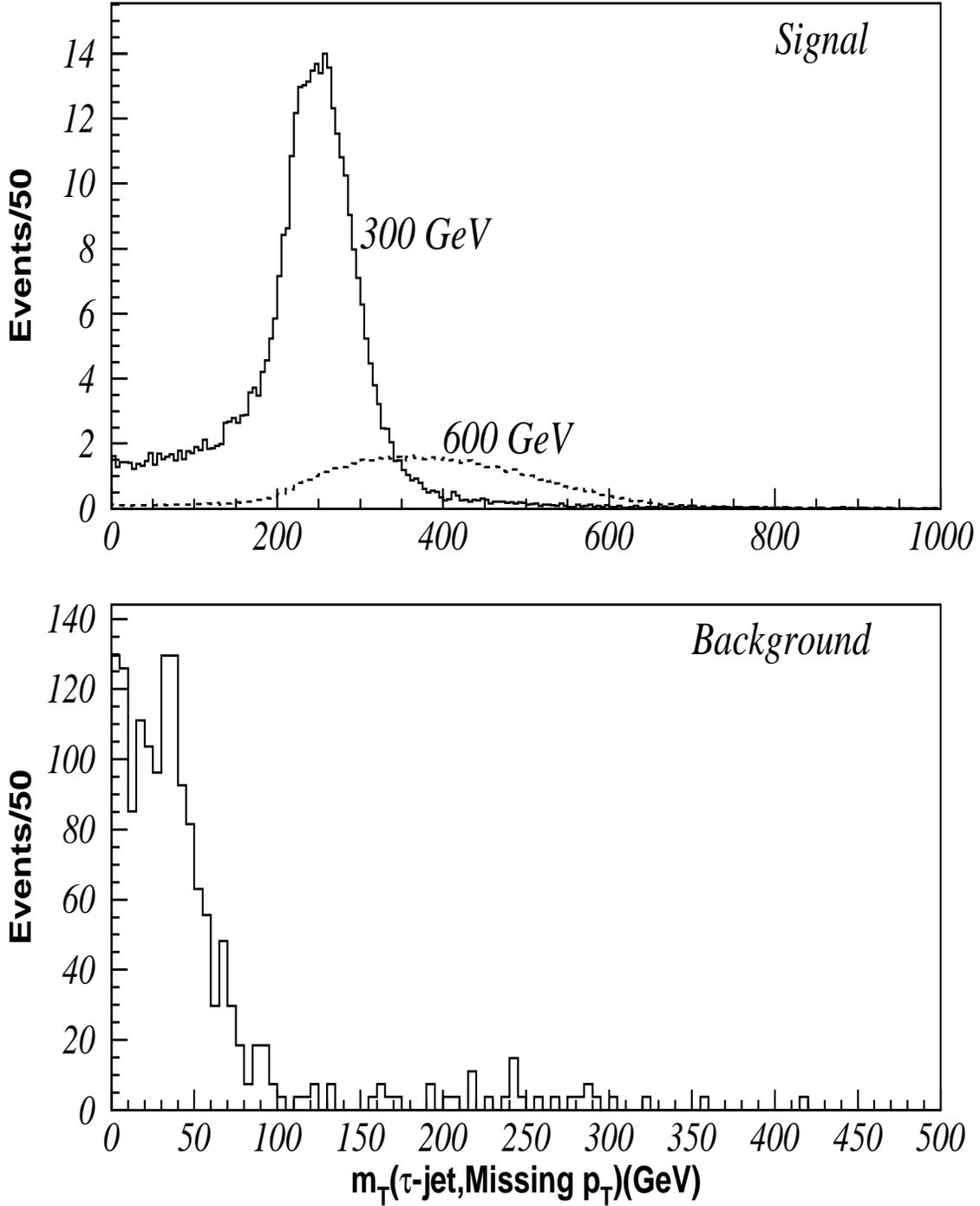}
\vskip-0.7cm
\caption{The number of events are shown against transeverse mass $m_T$ 
for signal and background for 1-prong decay channel of
$\tau$-jets. These are subject to
$p_T^{\tau-jet}> $100 GeV, $R_{1}>$0.8 and $E{\!\!\!/}_T >$ 100 GeV.
The masses of charged Higgs and $\tan\beta$ are same as in previous figures.
}
\end{figure}

\begin{figure}
\vskip-1cm
\includegraphics[width=6.5in, height=8.3in]{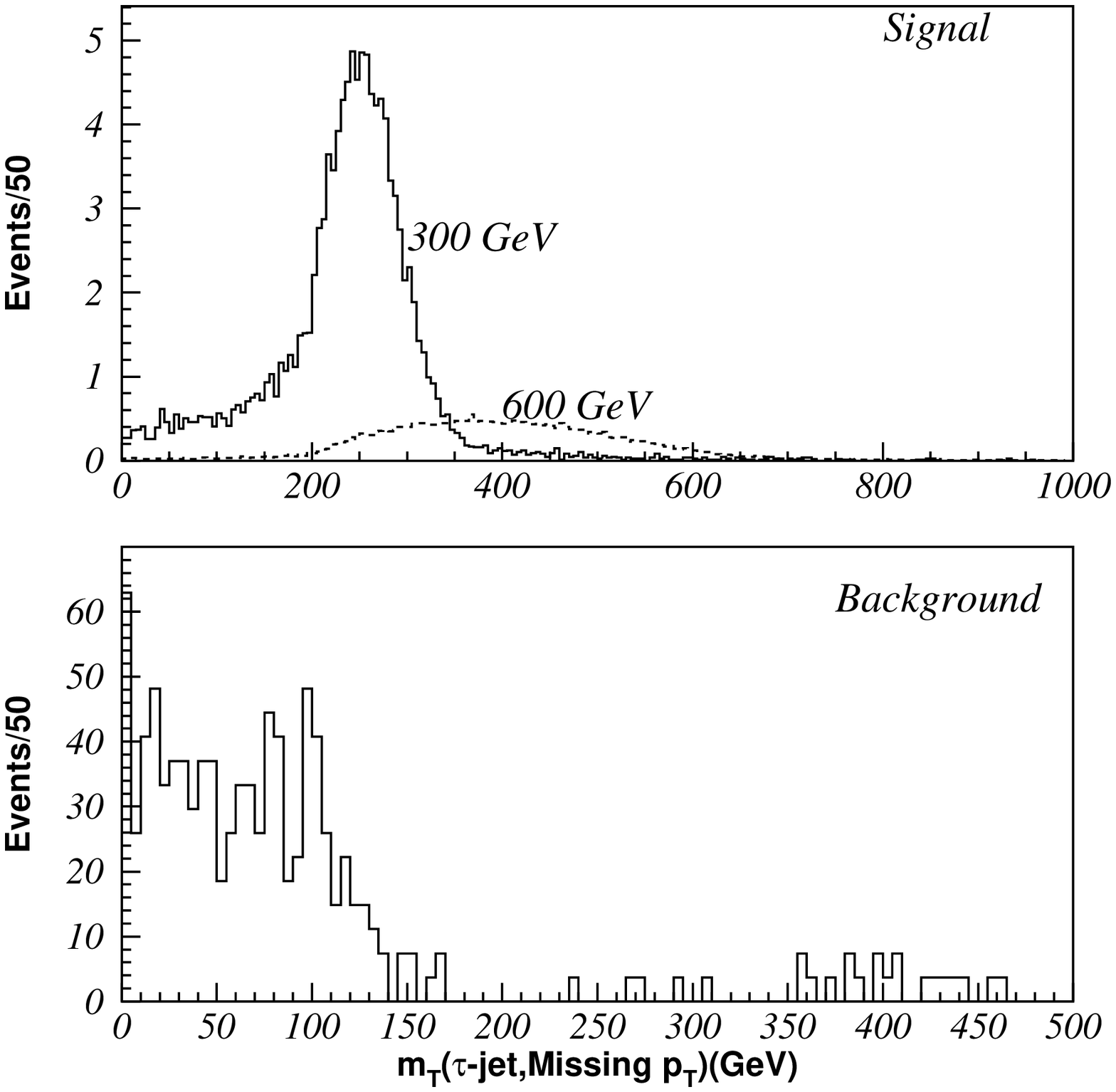}
\vskip-0.7cm
\caption{Same as in Fig.5, but for 3-prong decay channel of $\tau$ jets.
}
\end{figure}

\begin{figure}[htbp]
\begin{center}
\vskip-5cm
\hskip-1cm\centerline{\epsfig{file=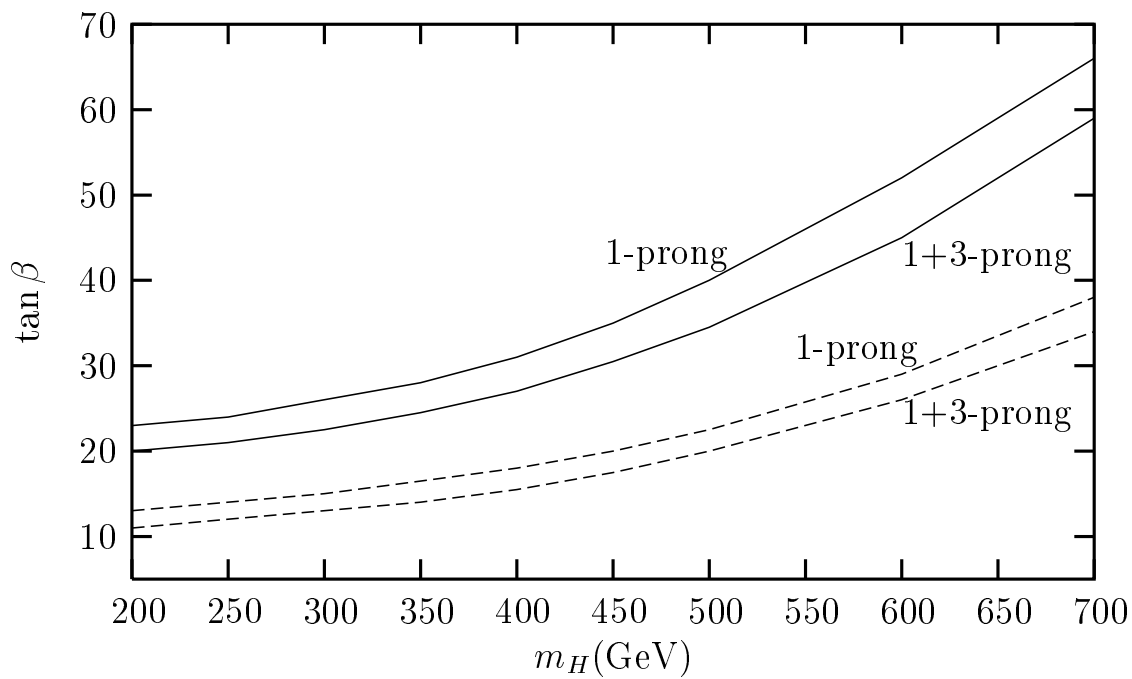,width=25cm}}
\vskip-15cm
\end{center}
\vskip-6cm
\caption
{
Discovery limits of charged Higgs are shown as functions of $\tan\beta$ for integrated  
luminosity ${\cal L}=30$ fb$^{-1}$(solid lines) and  
${\cal L}=100$ fb$^{-1}$(dashed lines) for $H^\pm \to \tau \nu$ where 
$\tau$ decays hadronically in both 1-prong and 3-prong channels. In each case
the contribution from 1-prong channel only are shown by upper lines where as 
contribution from the combined 1 and 3-prong channels are shown by lower lines.
}
\end{figure}

\end{document}